\documentclass[doublecol]{epl2} 
\usepackage{amssymb}
\newcommand{\half}{\frac{1}{2}}

\newcommand{\sxth}{\frac{1}{6}}
\title{Dark galactic halos without dark matter}
\author{R. K. Nesbet}
\institute{
IBM Almaden Research Center,
650 Harry Road,
San Jose, CA 95120-6099, USA 
}
\abstract{
Using standard Einstein theory, baryonic mass cannot account for 
observed galactic rotation velocities and gravitational lensing, 
attributed to galactic dark matter halos.  
In contrast, theory constrained by Weyl conformal scaling
symmetry explains observed galactic rotation in the halo region without 
invoking dark matter.  An explanation of dark halos, gravitational 
lensing, and structural stabilization, without dark matter and 
consistent with conformal theory, is proposed here.  Condensation of 
uniform primordial matter into a material cloud or galaxy vacates a 
large surrounding spherical halo.  Within such an extended vacancy in 
the original cosmic background mass-energy density, conformal theory 
predicts centripetal acceleration of the observed magnitude.
}
\begin{document}
\maketitle
%\begin{center} For {\em Europhysics Letters} \end{center}
%\keywords{galactic halos \and dark matter}
% body of paper here
%***************** VM screen width ************************************%
\section{Introduction}
\par The concept of dark matter galactic halos originates from 
dynamical studies\cite{OAP73,OPY74} which indicate that a spiral galaxy
such as our own would lack long-term stability if not supported by an
additional gravitational field from some unseen source. This concept is
supported by other cosmological data\cite{SAN10}, including 
gravitational lensing\cite{WCW79,HFK98}.  Excess centripetal 
acceleration observed in orbital velocities and lensing is attributed 
in standard $\Lambda$CDM theory\cite{DOD03,SAN10,SAL07} to a dark
matter halo.  A central conclusion of $\Lambda$CDM cosmology is that
inferred dark matter outweighs baryonic matter.  
\par In the $\Lambda$CDM model, an isolated galaxy is considered to be
surrounded by a much larger spherical dark matter halo.  
What is actually observed is a halo of gravitational field that
deflects photons in gravitational lensing and increases the velocity
of orbiting mass particles.  It will be shown here that conformal 
theory\cite{MAN06,MAN12,NES13}, which modifies both Einstein-Hilbert
general relativity and the Higgs scalar field model, supports 
an alternative interpretation of lensing and anomalous rotation 
as gravitational effects due to depletion of the cosmic background 
by concentration of diffuse primordial mass into an observed galaxy.
\par In standard theory, Poisson's equation determines a distribution
of dark matter for any unexplained gravitational field.  If dark 
matter interacts only through gravity, the concept of dark matter
provides a compact description of observed phenomena, not a falsifiable
explanation.  Postulated universal conformal symmetry\cite{NES13} 
promises a falsifiable alternative, not requiring any novel elementary 
fields.
\par Conformal gravity theory is based on the conformally invariant Weyl
tensor\cite{WEY18}. Possible formal difficulties\cite{FLA06,BAV09,YOO13}
have been discussed and resolved in detail\cite{MAN07,MAN11,NESM5}.  
Conformal Higgs theory is a uniquely defined version of standard 
$\phi^4$ field theory\cite{MAN06,NESM1}.

\par As a galaxy forms, background matter density $\rho_m$ condenses
into observed galactic density $\rho_g$.  Conservation of mass and
energy requires total galactic mass $M$ to be missing from a depleted
background.  Since the primordial density is uniform and isotropic, the
depleted background can be modeled by an empty sphere of radius $r_H$, 
such that $4\pi\rho_m r_H^3/3=M$.  In particular, the integral of 
$\rho_g-\rho_m$ must vanish.  The resulting gravitational effect, 
derived here from conformal theory, describes a dark halo.   
This halo model accounts for the otherwise remarkable fact that 
galaxies of all shapes are embedded in essentially spherical halos. 
Conformal theory relates the radial acceleration parameter of anomalous
galactic rotation to the observed acceleration of Hubble expansion. 
\par Assuming that galactic mass is concentrated within an average 
radius $r_g$, the ratio of radii $r_H/r_g$ should be very large, the 
cube root of the mass-density ratio $\rho_g/\rho_m$.  Thus if the latter
ratio is $10^5$ a galaxy of radius 10kpc would be accompanied by a halo 
of radius $10\times 10^{5/3}=464$kpc.  Equivalence of galactic and 
displaced halo mass resolves the paradox for $\Lambda$CDM that despite 
any interaction other than gravity, the amount of dark matter inferred 
for a galactic halo is strongly correlated with the galactic luminosity 
or baryonic mass\cite{SAL07,MCG05}.
\par Together with conformal theory of anomalous galactic rotation
velocities\cite{MAN06,MAN12} and Hubble expansion\cite{NESM1},
the present halo model removes any need to invoke dark matter for
major isolated galactic phenomena\cite{NES13}.  This suggests 
that when applied to other cosmological phenomena, galactic growth
and interaction and the original big-bang itself, conformal
theory will provide an alternative paradigm to $\Lambda$CDM. 
It is shown here that the Newton-Einstein model of galactic
growth and interaction is substantially altered by conformal theory.

\section{Gravitational effect of a depleted halo}
\par What, if any,  would be the gravitational effect of a depleted 
background density?  An analogy, in well-known physics, is impurity 
scattering of electrons in conductors.  In a complex material with
a regular periodic lattice, independent electron waves are by no means 
trivial functions, but they propagate without contributing to scattering
or resistivity unless there is some lattice irregularity, such as a
vacancy or substituted atom.  Impurity scattering depends on the
difference between impurity and host atomic T-matrices\cite{NES98}.
Similarly, a photon or isolated mass particle follows a geodesic in the
cosmic background unless there is some disturbance of uniform 
background density $\rho_m$.  
\par A gravitational halo is treated here as a natural consequence of
galaxy formation.  Condensation of matter from the primordial uniform
mass-energy distribution leaves a depleted sphere that has an explicit
gravitational effect. The implied subtracted mass, which integrates to 
minus the total galactic mass, cannot be ignored.  If this mass were 
simply removed, the analogy to vacancy scattering in solids implies a 
lensing effect.  A background geodesic is no longer a geodesic 
in the empty sphere.
\par Following the basic concepts of general relativity, deflection of
a geodesic would be observed as radial acceleration of orbiting mass
particles. It will be shown here that conformal 
theory\cite{MAN06,NESM1,NES13} explains the observed centripetal 
character of halo gravitation, without invoking dark matter.
In standard theory, subtracted density $-\rho_m$ would predict
centrifugal acceleration, contrary to observation. 
\par Analyses of galactic rotation velocities using
conformal gravity\cite{MAN06,MAN97,MAO11,MAO12,OAM12}, the MOND 
model\cite{MCG11,FAM12},
or the scalar-vector-tensor theory of 
Moffat\cite{MOF06,BAM06}, all without dark matter, fit observed data.
Analysis of Hubble expansion using the conformal Higgs scalar field 
model\cite{NESM1,NESM2} agrees with consensus cosmological 
data\cite{KOM09,KOM11} back to the CMB epoch, also without assuming 
dark matter.  
\par The fit of conformal gravity to anomalous galactic rotation data
implies a significant effect of the cosmic background, external to a 
baryonic galactic core\cite{MAK89,MAN97,MAN06}.  The inferred 
centripetal radial acceleration, independent of galactic structure
and mass, is attributed here to a galactic halo depleted of its
original mass density.  This is consistent with conformal 
theory of both galactic rotation and Hubble expansion\cite{NES13}.
Observed centrifugal acceleration of Hubble expansion\cite{RAS98,PER99}
is verified by the conformal Higgs model\cite{NESM1}.  As will be shown 
here, the change of Hubble acceleration due to depleted background mass 
density is consistent with the observed centripetal acceleration due to 
a gravitational halo.

\section{Summary of relevant theory}
\par Variational theory for fields in general relativity is a 
straightforward generalization of classical field 
theory\cite{WEI72,MAN06,NES03}.
Given Riemannian scalar Lagrangian density ${\cal L}$, action integral
$I=\int d^4x \sqrt{-g} {\cal L}$ is required to be stationary for
all differentiable field variations, subject to appropriate boundary 
conditions. $g$ here is the determinant of metric tensor $g_{\mu\nu}$.
Gravitational field equations are determined by metric functional 
derivative
$X^{\mu\nu}= \frac{1}{\sqrt{-g}}\frac{\delta I}{\delta g_{\mu\nu}}$.
Any scalar ${\cal L}_a$ determines energy-momentum tensor 
$\Theta_a^{\mu\nu}=-2X_a^{\mu\nu}$,
evaluated for a solution of the field equations.
\par Strict conformal symmetry\cite{WEY18,MAN06} requires invariance of 
field action integrals under local Weyl scaling, such that
$g_{\mu\nu}(x)\to g_{\mu\nu}(x)\alpha^2(x)$, where $\alpha(x)$ is real
and differentiable.
For a scalar field, $\Phi(x)\to\Phi(x)\alpha^{-1}(x)$.  A conformal 
energy-momentum tensor must be traceless.  This is true for massless 
fermion and gauge boson fields, but not for Einstein tensor 
$G^{\mu\nu}$.  Conformal theory\cite{MAN06,NES13}, which determines 
unique Lagrangian densities ${\cal L}_g$ for the 
metric tensor and ${\cal L}_\Phi$ for a scalar field, removes this 
inconsistency.  Subgalactic phenomenology is preserved, but gravitation 
on a galactic scale is modified\cite{MAK89}.  The conformal Higgs model
determines dark energy\cite{NESM1,NESM2}.
\par The conformal gravitational field equation is 
\begin{eqnarray} \label{cfeq}
X_g^{\mu\nu}+X_\Phi^{\mu\nu}=\half\Theta_m^{\mu\nu},
\end{eqnarray}
where index m refers to matter and radiation.   
An exact solution inside the halo radius is given by
\begin{eqnarray} \label{geq}
X_g^{\mu\nu}=\half\Theta_m^{\mu\nu}(\rho_g),
X_\Phi^{\mu\nu}=0, r\leq r_H.
\end{eqnarray}
The exact source-free solution\cite{MAK89} of the $X_g$ equation
is valid in the external halo, $r_g\leq r\leq r_H$, because $\rho_g$ 
vanishes.  Constants of integration fitted at $r_g$ and $r_H$ 
determine radial acceleration in this external halo.
$X_\Phi^{\mu\nu}=0$ is solved exactly using the modified 
Friedmann equation\cite{NESM1} with mass-energy density omitted.
%%%%##
\par Given any uniform mass-energy density $\rho$, for $r\leq r_H$,
field equation $X_\Phi^{\mu\nu}(\rho)=\half\Theta_m^{\mu\nu}(\rho)$ 
implies a modified Friedmann equation\cite{NESM1}.  Solution for scale 
factor $a(t)$ determines dimensionless Friedmann acceleration weight 
$\Omega_q(\rho)=\frac{{\ddot a}a}{{\dot a}^2}$\cite{NESM1}. For uniform 
$\rho=0, 0\leq r\leq r_H$, this solves the second of Eqs.(\ref{geq})
within the halo radius.  
However, the $X_\Phi$ equation includes dark energy, present regardless
of density $\rho$.  This produces the centrifugal acceleration
of background Hubble expansion\cite{NESM1,NES13}, which must be 
subtracted off in order to compute observable radial acceleration.
Observed geodesics, whose bending determines the extragalactic
centripetal acceleration responsible for lensing and orbital rotation
velocities, are defined relative to the cosmic background.
%%%%##
\par What is proposed here is to solve modified 
Friedmann equations for both $X_\Phi^{\mu\nu}(\rho_m)$ and 
$X_\Phi^{\mu\nu}(0)$ to obtain Friedmann acceleration weights at 
present time $t_0$, $\Omega_q(\rho_m)$ and $\Omega_q(0)$ respectively.
The observed radial acceleration is $\Delta\Omega_q=
\Omega_q(0)-\Omega_q(\rho_m)$, which cancels acceleration due to dark 
energy, much greater at present time than that due to baryonic mass.  
This implements the argument given above, that observed effects are 
determined by subtracting out the uniform background mass-energy  
density.  The gravitational field due to $\rho_g$ is augmented
by a halo field due to $-\rho_m$.
%%%%##

\section{The gravitational field within a depleted halo}
\par Analysis here is greatly simplified by solving the field equations
in two different metric systems, made compatible by choice of constants
of integration.  The exterior Schwarzschild (ES) metric is valid for
$X_g^{\mu\nu}$ and galactic rotational velocities in the external 
halo\cite{MAN06}, while the Robertson-Walker (RW) metric is valid for 
$X_\Phi^{\mu\nu}$ and Hubble expansion\cite{NESM1}.  
%%%%##
\par In conformal theory the most general metric outside a static 
spherically symmetric source density takes the exterior Schwarzschild
form\cite{MAK89}
\begin{eqnarray} 
ds^2_{ES}=-B(r)dt^2+\frac{dr^2}{B(r)}+r^2d\omega^2.
\end{eqnarray}
Here $c=\hbar=1$ and $d\omega^2=d\theta^2+\sin^2\theta d\phi^2$.
Because the Weyl tensor vanishes identically for the assumed uniform, 
isotropic void outside a spherical baryonic galactic core, conformal 
$X_g^{\mu\nu}=0$ in the external halo.  An exact solution of this 
source-free field equation outside source radius $r_g$ is given in the
ES metric by potential function\cite{MAK89}
\begin{eqnarray}
B(r)=1-2\beta/r+\gamma r-\kappa r^2, r\geq r_g.
\end{eqnarray}
A circular orbit outside $r_g$ is stable for velocity such that
$v^2=\half r\frac{dB}{dr}=\beta/r+\half \gamma r-\kappa r^2$.
Centripetal radial acceleration is $v^2/r$.
An assumed exact solution of the inhomogeneous field equation 
\begin{eqnarray}
X_g^{\mu\nu}(\rho_g)=\half\Theta_m^{\mu\nu}(\rho_g), r\leq r_g,
\end{eqnarray}
is extended out to halo radius $r_H$ by matching to this known 
external solution at $r_g$\cite{MAK89}.
\par In the depleted halo, integration parameters in 
$X_g^{\mu\nu}(\rho_g)=0$ are determined by continuity of the implied 
acceleration field at $r=r_g$ and at $r=r_H$, where acceleration
must vanish for continuity with the unmodified external cosmos.
Newtonian parameter $\beta=GM$ is proportional to total galactic 
mass $M$\cite{MAK89}.  Neglecting parameter $\kappa$, 
Mannheim\cite{MAN97} determined two universal parameters such 
that $\gamma=\gamma^*N^*+\gamma_0$ fits rotational data for eleven 
typical galaxies, not invoking dark matter. $N^*$ here is total visible 
plus gaseous mass in units of solar mass. This has recently been 
extended to 138 galaxies whose orbital velocities are known outside the 
optical disk\cite{MAO11,MAO12,OAM12}.  Parameter $\kappa$ was fitted 
as a global constant, not determined by a specific boundary condition. 
As discussed below, the present halo model indicates that $\kappa$
should be treated as a constant of integration whose value is
determined by physical boundary conditions, in particular at the
halo radius $r_H$. 
\par $\gamma_0$ is found to have a  universal value, independent of any 
particular galaxy.  It is attributed here to the subtracted density 
$\rho_m$ of the present halo model. 
In the Schwarzschild metric, a solution of homogeneous equation 
$X_g^{\mu\nu}=0$ valid at the galactic center excludes the 
singular potential $-2\beta_0/r$.  Extending out to halo radius $r_H$,
$B_0(r)=1+\gamma_0 r-\kappa_0 r^2$, where $\kappa_0=\gamma_0/2r_H$
terminates radial acceleration at the halo boundary.  There is no 
interior boundary value to determine acceleration parameter $\gamma_0$, 
which must however be consistent with the modified Friedmann equation 
implied by $X_\Phi^{\mu\nu}=0$ in the uniform, isotropic
Robertson-Walker metric, as shown below.  Radial acceleration due to 
$\rho_m$ is determined by its differential effect on this homogeneous 
field equation.  This follows from the fact that 
the mass-energy weight parameter in the modified Friedmann equation 
is much smaller than the dark energy weight, which determines Hubble 
expansion acceleration in the current epoch\cite{NESM1}.
\par A uniform, isotropic cosmos with Hubble expansion is described
by the Robertson-Walker (RW) metric\cite{WEI72}
\begin{eqnarray}
ds^2_{RW}=-dt^2+a^2(t)(\frac{dr^2}{1-kr^2}+r^2d\omega^2),
\end{eqnarray} 
for curvature parameter $k$.  The Weyl tensor and resulting conformal
Lagrangian density ${\cal L}_g$ vanish identically in this metric.
Lagrangian density ${\cal L}_\Phi$ of the conformal Higgs 
model\cite{NESM1} contains Higgs tachyonic mass term $w^2\Phi^\dag\Phi$ 
and $-\sxth R\Phi^\dag\Phi$, dependent on gravitational Ricci scalar 
$R$.  The scalar field equation has an exact solution such that 
$\Phi^\dag\Phi=\phi^2_0$, a spacetime constant if the time variation of
$R$ (on a cosmological time scale) is neglected.
The conformal Higgs model\cite{NESM1} determines a modified Friedmann 
cosmic evolution equation for scale parameter $a(t)$ that fits 
cosmological data back to the CMB epoch\cite{KOM09,KOM11} without 
invoking dark matter.  Higgs parameter $w^2$ becomes dark    
energy\cite{NESM1,NESM2} in this equation.  
\par The modified Friedmann equation\cite{NESM1,NES13} is
\begin{eqnarray}
 \frac{{\dot a}^2}{a^2}+\frac{k}{a^2}-\frac{\ddot a}{a}
    =\frac{2}{3}({\bar\tau}\rho+{\bar\Lambda}) ,
\end{eqnarray}
where energy density $\rho=\Theta_m^{00}$.
The parameters here are ${\bar\Lambda}=\frac{3}{2}w^2$ and 
${\bar\tau}=-3y^2/\phi_0^2$.  Numerical factor $y^2$, which must be 
determined from empirical data, allows for a dimensionless coefficient
of conformal ${\cal L}_\Phi$, nominally taken to be unity.  The Newton 
gravitational constant is replaced by a parameter of different sign
and magnitude.  The Einstein tensor is replaced by a traceless conformal
tensor.  Vanishing total trace\cite{MAN06} reduces the second Friedmann 
equation of standard theory to an identity.
\par Dividing by $({\dot a}/a)^2$ determines dimensionless sum rule
$\Omega_m+\Omega_\Lambda+\Omega_k+\Omega_q=1$ for the modified equation.
Radiation energy density is included in $\Omega_m$ here.
The dimensionless weights are $\Omega_m(t)$ for mass density, 
$\Omega_k(t)$ for curvature, and $\Omega_\Lambda(t)$ for dark energy, 
augmented by acceleration weight
$\Omega_q(t)=\frac{{\ddot a}a}{{\dot a}^2}$\cite{NESM1}.
Galactic rotation velocities, observed at relatively small redshifts,
determine radial acceleration values which can be compared with 
acceleration weights inferred from Hubble expansion data.
\vspace{0.5cm}
\section{Parameters $\gamma$ and $\kappa$}
\par Mannheim parameter $\gamma_0>0$, independent of galactic mass 
and structure, implies centripetal acceleration due to an isotropic
cosmological source\cite{MAN97}.
The parametrized gravitational field forms a spherical halo\cite{NES13}.
The depleted halo model removes a particular conceptual problem
in fitting $B(r)$ parameters $\gamma,\kappa$ of $ds^2_{ES}$ to 
galactic rotation data\cite{MAN97,MAN06,MAO11}.  In empirical parameter
$\gamma=\gamma^*N^*+\gamma_0$, $\gamma_0$ does not depend on
galactic mass, so must be due to the surrounding cosmos\cite{MAN97}.
Mannheim considers this to represent the net effect of distant matter,
integrated out to infinity\cite{MAN06}.  Since the interior term, 
coefficient $\gamma^*N^*$, is centripetal, one might expect the 
term in $\gamma_0$ to be centrifugal, describing attraction to 
an exterior source.  However, if coefficient $\gamma_0$ is due to a
depleted halo, the implied sign change determines net centripetal 
acceleration, in agreement with observation.  
\par Integration parameter $\kappa$, included in fitting rotation
data\cite{MAO11,MAO12}, cuts off gravitational acceleration at a
boundary radius.  In the halo model, $\kappa$ is determined 
by the boundary condition of continuous acceleration field at halo
radius $r_H$, determined by galactic mass, except for the nonclassical
linear potential term due to to the baryonic galactic core.
If this were determined directly by the $X_g$ equation for the material
galaxy, there is no obvious reason why it should terminate at the
halo boundary.  In a galactic cluster, mass conservation prevents 
interpenetrating halos, which may determine effective radii.
\par Universal parameter $\gamma_0$ implies a halo contribution to
function $v^2(r)$ that is the same for all galaxies, deviating only
as $r$ approaches the halo radius.  For galaxies of the same mass,
the full parameter $\gamma=\gamma^*N^*+\gamma_0$ implies identical
"dark matter" rotation curves, as exemplified for galaxies NGC2403
and UGC128 in Fig.1 of Reference\cite{MCG05}.  This supports the
empirical argument for a universal nonclassical rotation 
curve\cite{SAL07} and for a fundamental relationship between observed 
baryonic mass and inferred dark mass\cite{MCG05}.  Such a relationship 
is an immediate consequence of the depleted halo model. 
\par The halo model defines cutoff parameter $\kappa_{core}=GM/r_H^3$ 
and nonclassical $\kappa_0=\gamma_0/2r_H$, which enforce continuity
by terminating the acceleration field at $r_H$.
For $r\leq r_H$ but outside baryonic density bound $r_g$, 
$v^2$ for galactic rotation is the sum of three independent terms:
\begin{eqnarray}
v^2_{core}=\frac{GM}{r}(1-r^3/r_H^3),\\
v^2_{halo}=\half\gamma_0r(1-r/r_H),\\
\label{vncl}
v^2_{ncl}=\half N^*\gamma^*r(1-r/r_*).
\end{eqnarray}
\par Recent analysis of the $X_g$ field equation in the Schwarzschild
metric\cite{NESM5} indicates that Mannheim parameter $\gamma^*$ is in
fact determined by the $X_\Phi$ equation, in analogy to $\gamma_0$.
This would imply that $\gamma^*r$ should also cut off at $r_H$.
In either case, parameter $\kappa$ invalidates the Newtonian virial 
theorem for galactic clusters.  Implications are discussed below. 
It would be informative to fit galactic rotation data\cite{MAO11,MAO12} 
using $\kappa=\kappa^*N^*+\kappa_0$.  Deviations of $\kappa$ from a 
universal constant might determine halo radii dependent on baryonic 
mass, while $\kappa^*$ would test whether or not $\gamma^*r$ is cut off 
at the halo radius.

\section {Parameters $\Omega_k$, $\Omega_m$, and $\gamma_0$}
\par In the external halo, for $r_g\leq r\leq r_H$, a solution of
$X_g^{\mu\nu}=0$ is determined by ES metric constants of
integration $\beta, \gamma, \kappa$, proportional to galactic mass,
fitted at $r_g$ to an interior solution for $\rho_g\neq0$\cite{MAK89}.
$X_g^{\mu\nu}=0, 0\leq r\leq r_H$ for primordial uniform 
background $\rho_m$, because the Weyl tensor vanishes.
The exact background solution, regular at the coordinate
origin, can only affect Mannheim parameters $\gamma$ and $\kappa$.  
Subtraction of the background can be expressed in terms of 
Friedmann acceleration weight parameters $\Omega_q$ in the RW metric.
Geodesic deflection in the halo is due to net acceleration 
$\Delta\Omega_q=\Omega_q(halo)-\Omega_q(cosmos)$,
related to $\Delta\rho=\rho_g-\rho_m=-\rho_m$ within the halo.  
%%%%##
ES parameter $\gamma_0$ must be consistent with RW acceleration
$\Delta\Omega_q$.
\par $\Delta\Omega_q$ can be approximated in the halo (where  
$X_g^{\mu\nu}=0$) by solving equation $X_\Phi^{\mu\nu}(halo)=0$ in the 
RW metric to determine Friedmann weight $\Omega_q(halo)$, and by solving
$X_\Phi^{\mu\nu}(cosmos)=\half\Theta_m^{\mu\nu}(\rho_m)$ to determine 
$\Omega_q(cosmos)$.  The modified Friedmann equation, with relevant
parameters, is exact in each case. From the modified Friedmann sum rule,
net acceleration weight
$\Omega_q(cosmos)=1-\Omega_\Lambda-\Omega_k-\Omega_m$.
If the halo is a true vacuum, in which both $\rho_m$ and curvature
parameter $k$ vanish, $\Omega_q(halo)=1-\Omega_\Lambda$, so that
$\Delta\Omega_q\simeq\Omega_k+\Omega_m$, neglecting any change in
$\Omega_\Lambda$.
\par Positive $\rho_m$ implies $\Omega_m<0$, because of negative 
coefficient ${\bar\tau}$. Curvature $k\geq0$, such that
$\Omega_k(cosmos)\leq0$, implies a metric singularity at large $r$ which
may be related to Hubble radius $c/H_0$.  For nonnegative cosmic
curvature, $\Delta\Omega_q<0$, implying centripetal acceleration and 
positive $\gamma_0$, consistent with observed anomalous galactic
rotation\cite{MAN97,MAN06,MAO11}.
In Hubble units such that $\hbar=c=1$ and Hubble function
${\dot a}/a=1$, at present time $t_0$,
for $\gamma_0=3.06\times 10^{-30}/cm$, deduced from galactic rotation
data\cite{MAN97,MAN06,MAO11}, and Hubble length unit
$\frac{c}{H_0}=1.314\times 10^{28}cm$, radial acceleration implied by
$\gamma_0$ is $-\half\gamma_0\frac{c}{H_0}=-0.0201$.  The corresponding 
acceleration weight $\Delta\Omega_q\simeq\Omega_k+\Omega_m=-0.0201$ is
consistent with empirical $\Omega_k=-0.0125\pm0.0065$\cite{KOM11}.
\par Empirical $\gamma_0$ may give the most accurate current
estimate of $\Omega_k+\Omega_m$.  Hubble expansion data for redshifts
$z\leq 1$ can be fitted with $\Omega_k+\Omega_m=0$\cite{NESM1}.
Accurate data for large redshifts and for the CMB is required for an 
independent value of this sum and for the individual Friedmann  
weights.  Further analysis of the cosmic curvature parameter and
of the time variation of Higgs model parameters is needed.
Consistency of lensing and anomalous rotation constrains both
$\Lambda$CDM and conformal models.  Accurate rotational and lensing 
data for the same galaxy would provide a quantitative test of theory.

\section{Galaxy formation and galactic clusters}
A depleted halo accompanies any condensation of the cosmic 
background.  The resulting centripetal gravitation stabilizes such 
condensations, so that concentration and stabilization are concurrent.  
It is unnecessary to assume an initial concentration of dark matter. 
\par A new rule must be valid for galactic collisions: the total empty
volume must remain equal to the total galactic mass divided by
the cosmic background mass density $\rho_m$.  Halos cannot overlap,   
but must distort the background density to preserve total empty volume. 
This may have the effect of limiting galactic growth in clusters.
Two halos in contact would remove background matter that might 
otherwise fall into either central galaxy.
\par The intergalactic potentials derived here terminate either at the
halo radius or at some radius to be determined by dynamic modeling. 
Hence the Newtonian virial theorem is not valid for clusters. Relative
kinetic energy of colliding galaxies would be partially converted 
to thermal energy of the redistributed cosmic background density.
Detailed modeling, starting with two colliding halos with their
central galaxies, is needed to estimate the net thermal energy 
transferred to intergalactic dust or gas in a galactic cluster.
The need for dark matter in galactic clusters should be
reexamined on the basis of such modeling. 
\par The rules for galaxy formation 
are modified.  Current dynamical models of galaxy
and galactic cluster formation should be revised to take into account
concurrent halo formation, absence of dark matter, and modified
intergalactic potentials.  It would be informative to test conformal    
theory using a revised dynamical model.  Conclusions of
standard dynamical models should be reexamined.
\par The MOND model\cite{SAN10,FAM12}, which has been applied 
successfully to a wide variety of cosmological phenomena, parametrizes 
an assumed failure of Newtonian dynamics for acceleration less than a 
universal constant.  Similarly,
conformal theory finds new gravitational effects (acceleration
parameters $\gamma$) when Keplerian radial acceleration in a galaxy   
drops to a value comparable to that due to the universal effect of a 
galactic halo, as considered here. This suggests that other successful
applications of MOND, not yet studied by conformal theory, will
also turn out to be explained when such studies are carried out.

\section{Conclusions}
\par  Galaxy formation by condensation from the primordial cosmos 
implies a gravitational halo field due to depletion of the original 
uniform isotropic mass distribution.  Standard Einstein-Hilbert theory, 
implying centrifugal acceleration of a photon or orbiting mass particle 
in a depleted halo, is contradicted by observed centripetal lensing 
and enhanced rotation velocities, implied by conformal theory.
\par Conformal gravity, the conformal Higgs model, and the present
depleted halo model are mutually consistent.  Together they account
for observed excessive galactic rotation
velocities, Hubble expansion, stabilization of growing galaxies, and
galactic lensing, without invoking dark matter, and explain the
source and magnitude of dark energy.
%***************** VM screen width ************************************%

\end{document}